\documentclass[prl,twocolumn,superscriptaddress]{revtex4-2}
\usepackage{amsmath}
\usepackage{amssymb}
\usepackage{color}
\usepackage{graphicx}
\usepackage[colorlinks,bookmarks=false,citecolor=blue,linkcolor=red,urlcolor=blue]{hyperref}

\begin{document}
\title{Universal and non-universal contributions of entanglement under different bipartitions}

\author{Zhe Wang}
\email{wangzhe90@westlake.edu.cn}
\affiliation{Department of Physics, School of Science and Research Center for Industries of the Future, Westlake University, Hangzhou 310030, China}
\affiliation{Institute of Natural Sciences, Westlake Institute for Advanced Study, Hangzhou 310024, China}

\author{Chunhao Guo}
\affiliation{Department of Physics, School of Science and Research Center for Industries of the Future, Westlake University, Hangzhou 310030, China}
\affiliation{Institute of Natural Sciences, Westlake Institute for Advanced Study, Hangzhou 310024, China}

\author{Bin-Bin Mao}
\affiliation{Department of Physics, School of Science and Research Center for Industries of the Future, Westlake University, Hangzhou 310030,  China}
\affiliation{School of Foundational Education, University of Health and Rehabilitation Sciences, Qingdao 266000, China}

\author{Zheng Yan}
\email{zhengyan@westlake.edu.cn}
\affiliation{Department of Physics, School of Science and Research Center for Industries of the Future, Westlake University, Hangzhou 310030, China}
\affiliation{Institute of Natural Sciences, Westlake Institute for Advanced Study, Hangzhou 310024, China}

\begin{abstract}
Entanglement entropy (EE) is a fundamental probe of quantum phases and critical phenomena, which was thought to reflect only bulk universality for a long time. Very recently, people realized that the microscopic geometry of the entanglement cut can induce distinct entanglement-edge modes, whose coupling to bulk critical fluctuations may alter the scaling of the EE. However, this perception is very qualitative and lacks quantitative consideration.
Here, we investigate this problem through high-precision quantum Monte Carlo simulations combined with the analysis of scaling theory to build a quantitative understanding. By considering three distinct bipartitions corresponding to three surface criticality types, we reveal a striking dependence of the constant term \(\gamma\) on the microscopic cut at the quantum critical point. Notably, cuts that generate extra gapless edge modes yield a sign reversal in \(\gamma\) compared to those producing gapped edges. We explain this behavior via a modified scaling form that incorporates contributions from both bulk and surface critical modes. Furthermore, we demonstrate that the derivative of EE robustly extracts the bulk critical point and exponent \(\nu\) regardless of the cut geometry, providing a reliable diagnostic of bulk universality in the presence of strong surface effects. Our work for the first time establishes a direct quantitative connection between surface criticality and entanglement scaling, challenging the conventional view that EE solely reflects bulk properties and offering a refined framework for interpreting entanglement in systems with boundary-sensitive criticality.
\end{abstract}

\date{\today}
\maketitle

\textit{\color{blue} Introduction.---} Entanglement entropy (EE) has become a fundamental diagnostic of quantum phases and critical phenomena, offering deep insights into the non-local correlations inherent in quantum many-body systems~\cite{Amico2008,Laflorencie2016}. For gapped ground states, the EE typically satisfies an "area law", scaling with the boundary size of the subsystem~\cite{Eisert2010}. More remarkably, deviations from this leading behavior often carry universal information. In one-dimensional (1D) critical systems governed by conformal field theory (CFT), the area law is often modified by a logarithmic correction, whose coefficient is proportional to the central charge~\cite{Holzhey1994,Vidal2003,Calabrese2004,Calabrese2009Entanglement,Hong2012Identifying}. In two-dimensional (2D) quantum critical points (QCPs), universal contributions are further expected in the form of logarithmic terms from sharp corners and constant terms in the EE for smooth bipartitions—both widely believed to reflect properties of the underlying bulk CFT~\cite{Calabrese2004,Casini2007Universal,Metlitski2009,Casini2009Entanglement,Kallin2011a,Fradkin2006Entanglement,Laflorencie2016,Bueno2016Bounds,Bueno2015Universal,Bueno2015Universality,Helmes2014,Helmes2016Universal}. However, a growing body of evidence suggests that this picture may be incomplete: the microscopic details of the entanglement cut, especially in systems with spatially anisotropic interactions, can significantly alter the scaling of the EE. It has caused a long-standing confusion.

Recent studies in dimerized magnets and deconfined critical systems have shown that the manner in which the entanglement boundary intersects the underlying interaction bonds can lead to distinct entanglement-edge excitations, which in turn couple to the original bulk critical fluctuations and modify entanglement scaling~\cite{zhao2022measuring,Zhao2022a,DEmidio2024,wang2025boundary,zhu2025Bipartite}.
What's more, there is a potential correspondence between the entanglement-edge and physical edge (real cut) to lead ``entanglement boundary criticality" phenomena affecting the EE behaviors \cite{zhu2025Bipartite}. For convenience, we use ``edge/boundary" directly to represent entanglement-edge/boundary in the following. In particular, the emergence of gapless edge modes—associated with special or extraordinary surface universality classes—has been suggested to leave a pronounced imprint on the entanglement spectrum (ES) and the universal part of the EE. Yet, a quantitative understanding of how surface criticality contributes to entanglement scaling, and whether its effects can be systematically disentangled from bulk universal behavior, remains an open and pressing challenge.  

In this work, we address this challenge through high-precision quantum Monte Carlo simulations~\cite{Sandvik1992a, Sandvik1999, Syljuasen2002, Sandvik2010a, Yan2019a, Yan2022} of dimerized spin-\(1/2\) Heisenberg antiferromagnets on the square lattice~\cite{Matsumoto2001a,Wenzel2008Evidence,Yasuda2013Monte,Ma2018Anomalous,Fritz2011Cubic,Jiang2012Monte,Ding2018Engineering,Wang2025Extraordinary}. We consider two dimerization patterns—columnar and staggered—and three distinct entanglement bipartitions that cut through different bond types (weak, strong, or staggered). Our results reveal that the constant term \(\gamma\) in the EE at the QCP depends strongly on the cut edge, even for a fixed bulk universality class. Strikingly, we find that cuts that produce gapless physical edge modes lead to a sign reversal in \(\gamma\), relative to cuts that yield gapped edges.  

We explain this behavior via a modified scaling form that explicitly incorporates contributions from both critical bulk and surface gapless modes. Furthermore, we demonstrate that the derivative of EE (DEE) robustly captures the bulk critical point and exponent \(\nu\), regardless of cut details, thus providing a reliable probe of bulk universality even in the presence of strong surface effects. Our work establishes a direct quantitative link between surface critical behavior and entanglement scaling, challenges the conventional view that EE solely reflects bulk properties, and offers a refined framework for interpreting entanglement measures in systems with boundary-sensitive criticality.

\begin{figure}[!tb]
\centering
\includegraphics[width=\columnwidth]{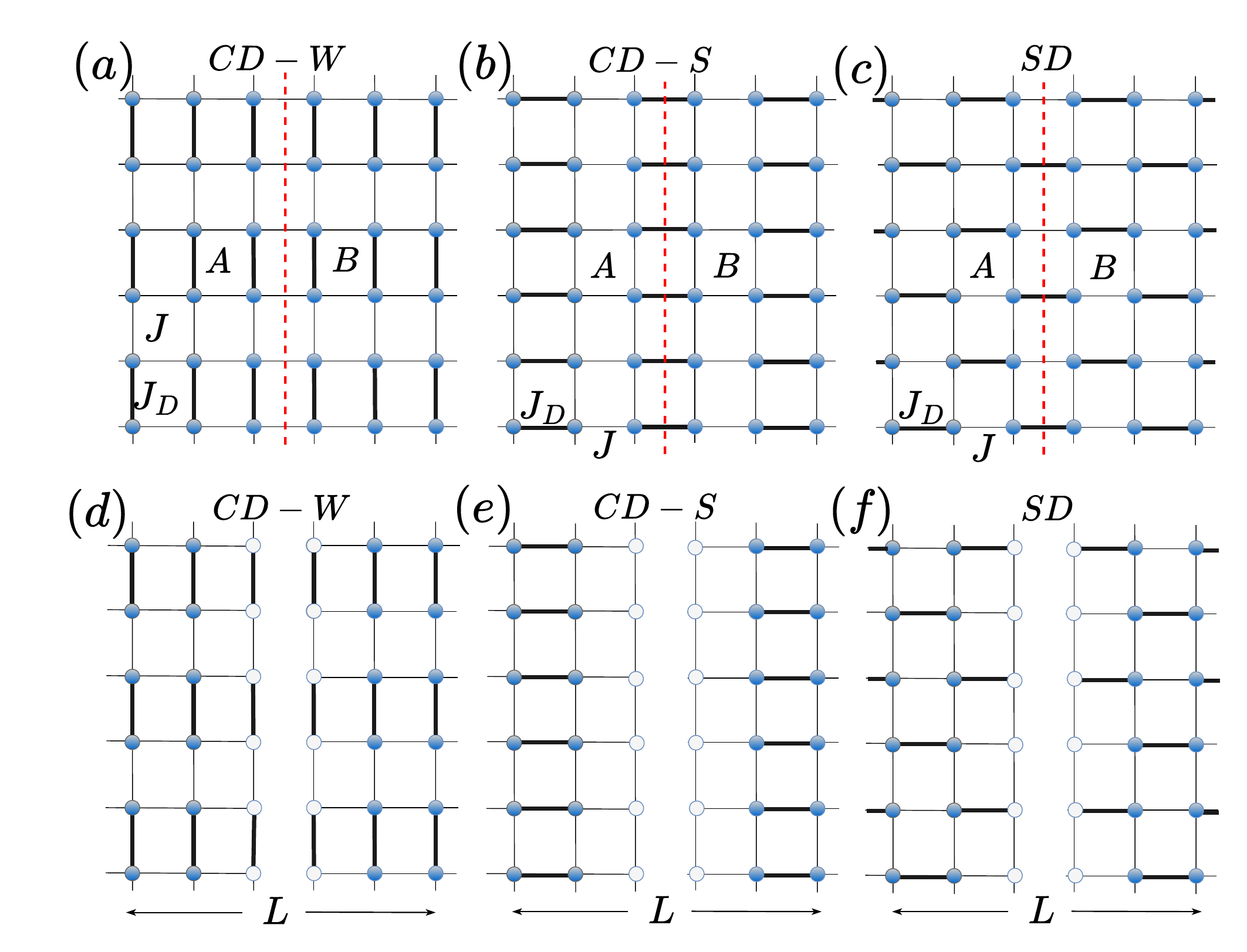}
\caption{ Columnar and staggered dimerized spin-1/2 Heisenberg models on the square lattice, with couplings $J$ and $J_D > J$ on alternating bonds. (a,d) Entanglement/physical boundary cuts weak bonds (CD-W).  (b,e) Cuts strong bonds (CD-S).  (c,f) Cuts staggered bonds in the staggered model (SD).  The system size is $L \times L$; subsystem $A$ is a $(L/2) \times L$ cylinder, separated from environment $B$ by dashed lines. Physical boundaries (open circles) denote true open edges.}

\label{fig:model}
\end{figure}

\textit{\color{blue} Models and method.---} In this work, we investigate the EE scalings of dimerized spin-1/2 Heisenberg antiferromagnets on the square lattice, as illustrated in Fig. \ref{fig:model}. The system is described by the Hamiltonian  
\begin{equation}
\mathcal{H} = J \sum_{\langle i,j \rangle} \mathbf{S}_i \cdot \mathbf{S}_j + J_{D} \sum_{\langle i,j \rangle'} \mathbf{S}_i \cdot \mathbf{S}_j,
\quad \text{(1)}
\end{equation} 
where \( J \) and \( J_{D} \) denote the exchange couplings on weak and strong bonds (represented by thin and thick lines, respectively). Two distinct dimerization patterns are considered: a columnar arrangement [Fig. \ref{fig:model}(a)] and a staggered configuration [Fig. \ref{fig:model}(c)], referred to as the columnar and staggered models, respectively.

When \( J_{D}/J \sim 1 \), both models exhibit long-range Néel antiferromagnetic order in the ground state. In the opposite limit \( J_{D}/J \gg 1 \), the ground state is well approximated by a product state of spin singlets formed on the strong bonds, resulting in a gapped, symmetry-preserving paramagnetic phase. Earlier numerical studies have established the existence of a continuous quantum phase transition between these two phases in both models~\cite{Matsumoto2001a,Wenzel2008Evidence,Yasuda2013Monte,Ma2018Anomalous,Fritz2011Cubic,Jiang2012Monte,Ding2018Engineering,Wang2025Extraordinary}. For the columnar model (staggered model), the QCP is located at \( J_{D}/J = 1.90951(4) \)~\cite{Matsumoto2001a,Yasuda2013Monte,Ma2018Anomalous} ( \( J_{D}/J = 2.5196(2) \)~\cite{Yasuda2013Monte,Ma2018Anomalous}), and extensive evidence confirms its description by the (2+1)D O(3) universality class~\cite{Matsumoto2001a,Yasuda2013Monte,Ma2018Anomalous,Ding2018Engineering}.

All our numerical simulations are carried out on $ L \times L $ square lattices with periodic boundary conditions imposed in both spatial directions. Given the dynamical critical exponent $ z = 1 $, which characterizes the system, we set the inverse temperature as $ \beta = 2L $ to ensure that the thermal length scale is comparable to the system size—thereby accessing the zero-temperature quantum critical regime.  The subsystem $ A $ is chosen as an $ (L/2) \times L $ cylindrical region, featuring two smooth boundaries parallel to one lattice direction. The total boundary length is thus $ \ell = 2L $. To compute the second R\'enyi EE, defined as $ S^{(2)} = -\ln \mathrm{tr}(\rho_A^2) $, where $ \rho_A $ is the reduced density matrix of subsystem $ A $, we employ stochastic series expansion (SSE) quantum Monte Carlo simulations~\cite{Sandvik1992a, Sandvik1999, Syljuasen2002, Sandvik2010a, Yan2019a, Yan2022}. The measurement is enhanced by a bipartite reweighting and annealing scheme~\cite{Wang2025Zhe,DEmidio2024}, which allows for efficient and accurate estimation of the EE in large-scale simulations.

\begin{figure}[htp]
\centering
\includegraphics[width=0.5\textwidth]{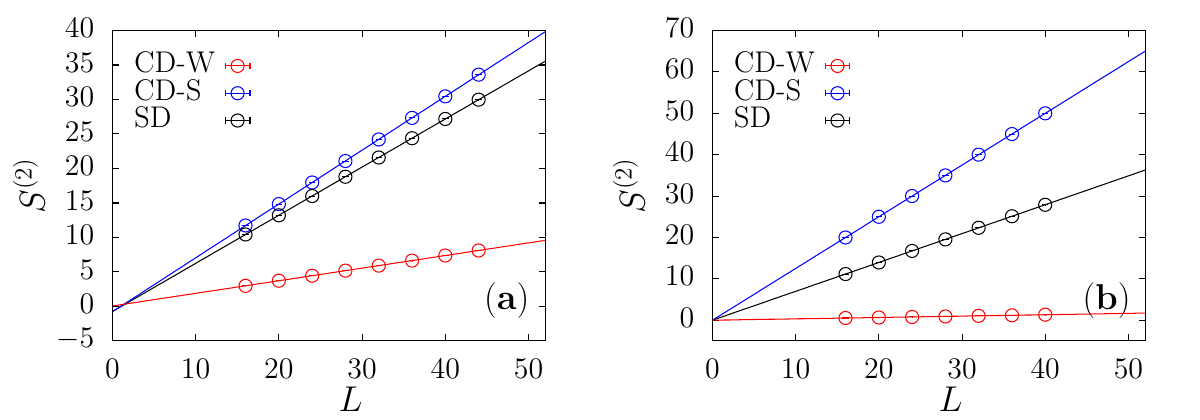}
\caption{Second Rényi EE $S^{(2)}$ under three bipartition schemes versus system size $L$, at the QCP $J_c$ (a) or in the gapped dimer phase at $J_c/2$ (b).  The fitting results are listed in Table \ref{ext1}. } 
\label{fig:ee}
\end{figure}

\textit{\color{blue} EE behaviors.---}From the perspective of CFT, the corner contributions to the EE, or the constant terms under smooth bipartitions (at fixed aspect ratio), typically reflect the universal properties of the underlying bulk critical point and do not depend on the microscopic details of how the subsystem is carved out~\cite{Calabrese2004,Casini2007Universal,Metlitski2009,Laflorencie2016}.

However, recent numerical simulations have revealed that the corner contributions to EE in the columnar dimerized Heisenberg model are consistent with field theory predictions for the (2+1)D O(3) critical point only when the entanglement bipartition boundary avoids all dimers~\cite{zhao2022measuring,Zhao2022a}. This highlights a strong dependence on the microscopic details of the cut. Furthermore, careful numerical and analytical studies near deconfined quantum critical points have provided evidence for qualitative differences between conventional square and tilted bipartitions, with results showing striking and dramatic effects~\cite{Deng2024,DEmidio2024}. Our recent work has demonstrated that the Li-Haldane-Poilblanc conjecture \cite{Li2008,poilblanc2010entanglement,liu2024b,li2024relevant}—the ES resembles the edge energy spectrum—remains approximately valid even when the bulk is gapless~\cite{zhu2025Bipartite}. This establishes, for the first time, a direct connection between surface critical behavior and entanglement properties. We find that edge modes associated with different surface universality classes significantly influence the low-energy structure of the entanglement spectrum, thereby altering the scaling of the entanglement entropy and preventing it from faithfully reflecting the bulk criticality of the system.

While these numerical results qualitatively indicate that boundary critical behavior indeed modifies the CFT predictions for entanglement, a pressing and important question remains: how, quantitatively, does boundary criticality contribute to or alter the scaling of the EE?

Corners along the entanglement boundary usually give rise to additional logarithmic corrections at QCP, which can obscure the contribution of edge modes. To isolate and more clearly probe the role of boundary criticality, we focus on a subsystem with a smooth (cornerless) entangling surface. In the quantum critical regime, the EE receives contributions from both short-range, non-singular correlations and long-range, singular fluctuations associated with bulk criticality. For a subsystem bounded by a smooth interface of length $2L$, the EE near the QCP takes the form~\cite{Metlitski2009}  
\begin{equation}
S(L, g) = a(g)\,L + \tilde{S}_0(x),
\label{eq:sell}
\end{equation}  
where $a(g)$ is a non-universal coefficient that depends analytically on the tuning parameter $g$. The second term, $\tilde{S}_0$, is a universal function of the dimensionless scaling variable $x = (g - g_c)\,L^{1/\nu}$~\footnote{In Ref.~\cite{Metlitski2009}, the universal scaling function was expressed as $S_0(\ell/\xi)$, with $\xi \propto |g - g_c|^{-\nu}$ being the correlation length. Here, we adopt $x = (g - g_c)\,\ell^{1/\nu} \propto (\ell/\xi)^{1/\nu}$ to facilitate a more transparent analysis of finite-size scaling.}.  

At the critical point ($g = g_c$), $\tilde{S}_0$ reduces to a universal constant $\tilde{S}_0(0)=\gamma$ determined solely by the universality class of the QCP and the aspect ratio of the subsystem. When the system is detuned away from criticality, $\tilde{S}_0(x)$ contributes to the extensive part of the EE. In the limit $|x| \gg 1$, this yields a correction scaling as $\tilde{S}_0(x) \simeq \gamma \,|g - g_c|^{\nu} L$.
However, our numerical results show that even for a fixed subsystem aspect ratio, the entanglement boundary's microscopic structure—specifically, how it intersects the underlying interacting bond pattern—has a significant impact on the universal scaling function $\tilde{S}_0(x)$. In particular, different bipartitions that cut through distinct types of interacting bonds yield markedly different values of $\tilde{S}_0(0) = \gamma$, despite all corresponding to the same bulk universality class.  To systematically investigate this effect, we consider three representative bipartition geometries:   the entanglement cut passes entirely through weak bonds (CD-W, see Fig. \ref{fig:model}(a)),  
the cut runs exclusively along strong bonds (CD-S, see Fig. \ref{fig:model}(b)),   the cut alternates between strong and weak bonds in a staggered fashion (SD, see Fig. \ref{fig:model}(c)).  

\begin{ruledtabular}
\begin{table}[!h]
\caption{ Fitting results for the data in Fig.~\ref{fig:ee}(a) ($J = J_c$) and Fig.~\ref{fig:ee}(b) ($J = J_c/2$), using the form $S^{(2)}(L) = aL + \gamma$. }
\begin{tabular}{l c c  c c }
 	  Cuts 	   	  & $a(J=J_c)$ 	   	&$\gamma (J=J_c)$ & $a(J=J_c/2)$ 	   	&$\gamma (J=J_c/2)$ \\
 	   	   	
\hline
    CD-W			& 0.183(1)   	&0.069(4)  & 0.033(1)   	&0.000(4)    \\
	CD-S			& 0.781(1)  	&-0.764(9)  & 1.2498(2)  	&0.000(3)      \\
    SD			& 0.698(1)   	&-0.739(7)   & 0.6977(2)   	&0.001(2)    \\

\end{tabular}
\label{ext1}
\end{table}
\end{ruledtabular}

As illustrated in the Fig.~\ref{fig:ee}(a) and Table~\ref{ext1}, the three distinct entanglement cuts yield different values of the $ \gamma $ at the QCP. Most strikingly, the signs of $ \gamma $ for CD-S and SD are opposite to that of CD-W. This difference provides a quantitative signature of how distinct surface critical behaviors contribute to entanglement scaling. The ordinary boundary provides a universal constant $ \gamma $ indentifying the bulk criticality \cite{zhu2025Bipartite,Metlitski2009,Metlitski2011}


To understand this, we first consider the physical edge states associated with each bipartition, under genuine open boundary conditions, as Fig.\ref{fig:model} (d-f) show. In the bulk-gapped phase away from criticality, the CD-W cut leaves behind dangling 1D dimerized chains  at the boundary (see Fig.~\ref{fig:model}(d)), which are themselves gapped~\cite{Ding2018Engineering,Wang2025Extraordinary}. Consequently, at the bulk critical point, the surface critical behavior is driven purely by bulk critical fluctuations, corresponding to the ordinary surface universality class.  In contrast, the CD-S and SD cuts (see Fig.~\ref{fig:model}(e) and (f)) give rise to effective one-dimensional chains: CD-S forms an antiferromagnetic spin-1/2 Heisenberg chain, while SD results in an effective ferrimagnetic chain—both of which are gapless~\cite{Ding2018Engineering,Wang2025Extraordinary}. These gapless edge modes couple to the bulk critical fluctuations, leading to nontrivially surface critical behaviors/modes. As a result, they belong to the special and extraordinary surface universality classes, respectively.

\begin{figure*}[!tb]
\centering
\includegraphics[width=\textwidth]{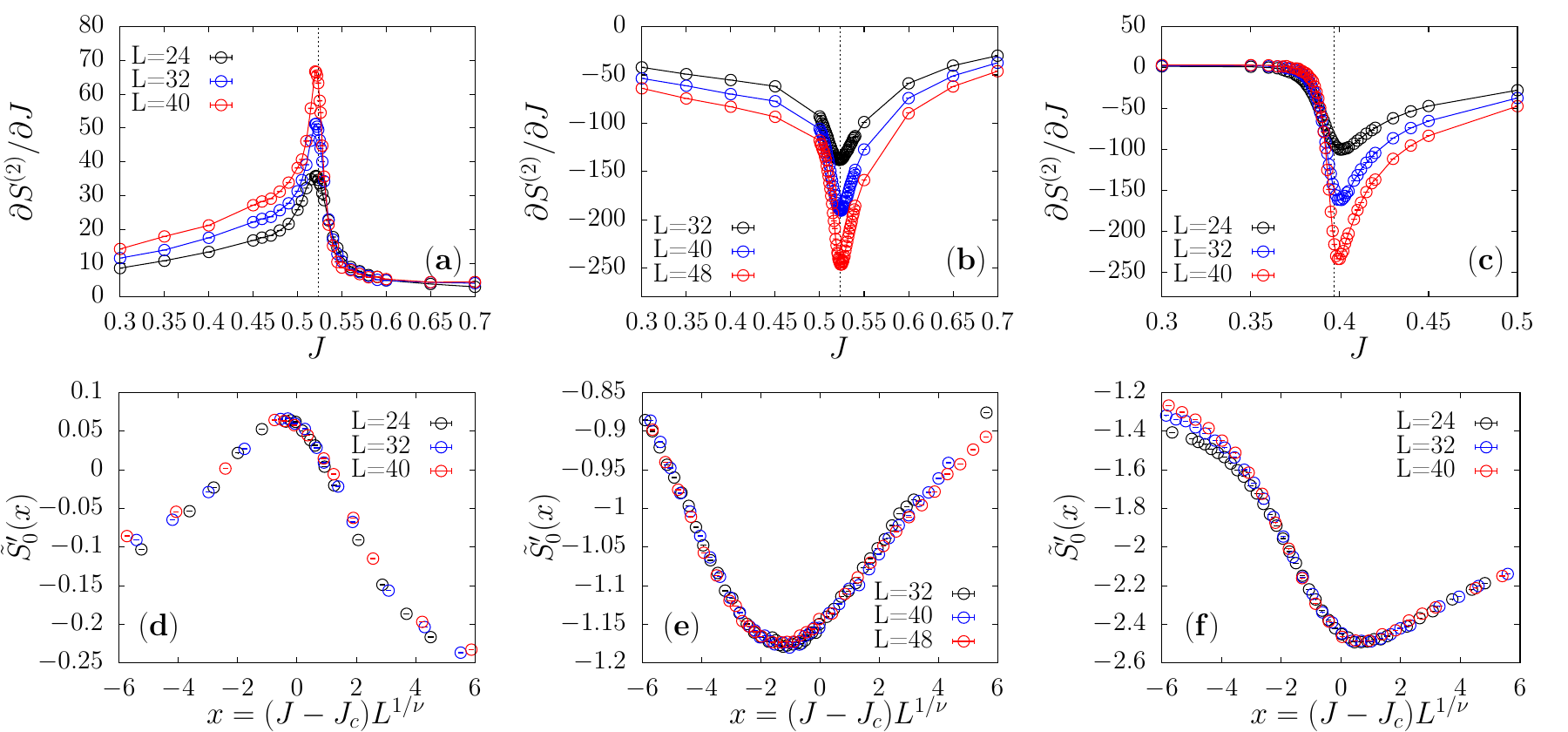}
\caption{The derivative of the second R\'enyi EE and the scaling function $\tilde{S}_{0}'(x)$ obtained from the data collapse analysis of the DEE near the transition.  (a,d) Data of the columnar model with  entanglement boundary cuts weak bonds (CD-W).  (b,e) Data of the columnar model with  entanglement boundary cuts strong bonds (CD-S). For columnar model $J_{D}=1.0$ is fixed, while $J$ is tuned around the QCP $J_{c}=0.52337(3)$.   (c,f) Data of the staggered model with  entanglement boundary cuts staggered bonds (SD). For staggered model, $J_{D}=1.0$ is fixed, while $J$ is tuned around the QCP $J_{c}=0.39692(1)$. }
\label{fig:qcp}
\end{figure*}

Then we correlate the boundary excitations with the ES. Remarkably, the Li-Haldane-Poilblanc conjecture \cite{Li2008,poilblanc2010entanglement}—that the ES resembles the edge energy spectrum—remains approximately valid even when the bulk is gapless~\cite{zhu2025Bipartite,Yan2023b,li2024relevant,liu2024b}. This correspondence allows us to interpret the entanglement structure in terms of physical edge excitations. Crucially, gapped edge modes do not significantly alter the low-energy ES; thus, the constant term $ \gamma $ in the EE captures only universal properties of the bulk criticality. In contrast, gapless edge modes actively modify the ES, contributing additional universal terms that encode the interplay between bulk and boundary criticality. These contributions are not only significant in magnitude but also appear with a sign opposite to that of the bulk term.

All in all, when gapless edge modes exist in the gapped bulk phase, their coupling to bulk criticality at the quantum phase transition leads to a modified scaling form for the EE:
\begin{equation}
S(L, g) = a(g)\,L + \tilde{S}_b(x) - \tilde{S}_s(x),
\label{eq:eebands}
\end{equation}
where $ \tilde{S}_b(0)=\gamma_b $ represents the universal contribution from the bulk criticality, and $ -\tilde{S}_s(x)=\gamma_s $ accounts for the negative correction arising from the nontrivial surface modes. This sign structure explains the observed reversal in the universal coefficient $  $ across different bipartitions and establishes a direct connection between surface universality classes and entanglement scaling.

Why is $ \tilde{S}_s(x) $ also a function of the dimensionless scaling variable $ x = (g - g_c)\,L^{1/\nu} $? As stated above, in the limit $ L/\xi \to 0 $, the system behaves as effectively critical, and the universal term $\tilde{S}_0(x)=\tilde{S}_b(x) - \tilde{S}_s(x) $ reduces to a constant $\gamma=\gamma_b-\gamma_s $. In the opposite limit $ L/\xi \to \infty $, the system obeys a pure area law for all three bipartition cuts (see Fig.~\ref{fig:ee}(b) and Table~\ref{ext1}).  Hence, $S =a(g)\frac{L^{d-1}}{a^{d-1}} + (\gamma_b-\gamma_s)\frac{L^{d-1}}{\xi^{d-1}}$. In the Eq.~(\ref{eq:eebands}), we adopt $x = (g - g_c)\,L^{1/\nu} \propto (L/\xi)^{1/\nu}$ to facilitate a more transparent analysis of finite-size scaling. We will provide further evidence for this by examining the derivative of the EE in the next section.


\textit{\color{blue} EE derivatives.---} EE has become a cornerstone for characterizing the universal and non-local properties of quantum many-body systems. In 2D QCPs, the corner contributions or constant terms in the EE are closely tied to the underlying CFT and are expected to encode universal information about the critical system~\cite{Calabrese2004,Casini2007Universal,Metlitski2009,Casini2009Entanglement,Kallin2011a,Fradkin2006Entanglement,Laflorencie2016,Bueno2016Bounds,Bueno2015Universal,Bueno2015Universality,Helmes2014,Helmes2016Universal}.  However, as our analysis above demonstrates, these contributions are not truly universal. They can be significantly influenced by boundary effects arising from how the entanglement cut intersects the microscopic interaction bonds—particularly through the emergence of edge critical modes. In models with broken spatial symmetries, such bond-dependent bipartition effects are unavoidable, leading to distinct entanglement signatures that depend on the specific geometry of the cut~\cite{DEmidio2024,zhu2025Bipartite}. This reveals a profound interplay between entanglement scaling and surface critical behavior, challenging the conventional view that entanglement measures solely reflect bulk universality.

A promising finding is that the derivative of the EE (DEE) always captures the universal properties of the bulk critical point. In our previous work, we derived the scaling form of the EE derivative~\cite{Wang2025Universal}:
\begin{equation}
\frac{\partial S(L,g)}{\partial g} = a'(g)L + \tilde{S}'_0(x) L^{1/\nu},
\label{eq:dell}
\end{equation}
where $ g $ denotes a generic parameter in the Hamiltonian. As shown in the Fig.~\ref{fig:qcp}, for all three bipartition—corresponding to different ways of cutting the interaction bonds—the DEE consistently captures the location of the bulk critical point and the critical exponent $ \nu $, thereby isolating the universal bulk information. Specifically, fitting the scaling relation Eq.~(\ref{eq:dell}) to the DEE data, we find $J_{c}=0.522(1)$ and $\nu=0.71(4)$ for CD-W cut, $J_{c}=0.527(2)$ and $\nu=0.71(1)$ for CD-S cut, $J_{c}=0.398(1)$ and $\nu=0.71(1)$ for SD cut. These results are consistent with the known QCPs and the critical exponents $\nu=0.7073$ for the
(2+1)D O(3) transition~\cite{ Matsumoto2001a,Nahum2015a}.

Beyond this robustness, two important observations deserve emphasis. First, we clearly observe that the peaks in the DEE for the CD-S and SD cuts diverge with an opposite sign compared to that of the CD-W cut. Since the singular behavior of the derivative arises from the universal scaling function in Eq.~(\ref{eq:eebands}), this sign reversal directly explains the opposite sign of the constant term $ \gamma $ at the critical point. Second, as analyzed earlier, this sign reversal in both the constant term and the DEE peak originates from the presence of gapless surface critical modes, which differ across the bipartition types. The fact that the DEE still accurately captures the bulk critical exponents despite the sign reversal in the peak provides further evidence that $ \tilde{S}_s(x) $—the surface contribution—is itself a function of the dimensionless scaling variable $ x = (g - g_c) L^{1/\nu} $. This confirms that surface criticality follows the same scaling framework as the bulk, and its effects are encoded in a controlled, universal manner within the entanglement response.

\textit{\color{blue} Conclusion and discussions.---}In summary, we have quantitatively investigated the interplay between bulk and surface criticality in shaping the scaling of entanglement entropy. By examining dimerized Heisenberg antiferromagnets with distinct entanglement bipartitions, we demonstrated that the supposedly universal constant term \(\gamma\) at the quantum critical point depends critically on the microscopic structure of the cut, even within the same bulk universality class. The sign reversal of \(\gamma\) observed for bipartitions producing gapless edge modes—as opposed to gapped ones—serves as a direct signature of how surface universality classes imprint themselves on entanglement measures.

These findings compel us to revise the conventional CFT-based expectation that smooth-bipartition contributions to the EE are purely bulk universal. Instead, we have shown that surface critical modes actively participate in the entanglement structure, leading to a modified scaling form that separates bulk and surface contributions. The robustness of the entanglement‑entropy derivative in extracting the bulk critical exponent \(\nu\), regardless of the cut geometry, further highlights that while the total EE encodes surface physics, its derivative remains a faithful probe of bulk universality—a crucial insight for future numerical and experimental studies.

Our work establishes a concrete and quantitative connection between boundary critical phenomena and entanglement scaling, offering a refined paradigm for interpreting entanglement in spatially anisotropic systems. This framework is expected to be relevant for a broad class of quantum materials and models where edge or surface modes are present, including deconfined quantum critical points, topological phases, and systems with engineered boundaries. Ultimately, our results underscore that entanglement entropy is not merely a bulk diagnostic, but a sensitive and rich observable that bridges bulk quantum criticality and surface physics.

\begin{acknowledgements}
\textit{\color{blue} Acknowledgements.---} We thank the helpful discussions with Long Zhang, Meng Cheng, Yan-Cheng Wang. ZW is supported by the China Postdoctoral Science Foundation under Grants No.2024M752898. ZL is supported by the China Postdoctoral Science Foundation under Grants No.2024M762935 and NSFC Special Fund for Theoretical Physics under Grants No.12447119. 
The work is supported by the Scientific Research Project (No.WU2025B011) and the Start-up Funding of Westlake University.
The authors thank the high-performance computing centers of Westlake University and the Beijng PARATERA Tech Co.,Ltd. for providing HPC resources.
\end{acknowledgements}

\end{document}